\documentclass[aps,floatfix,prd,nofootinbib,superscriptaddress,reprint,showpacs,10pt,preprintnumbers,longbibliography]{revtex4-1}
\usepackage[utf8]{inputenc}
\usepackage{graphicx}
\usepackage{float}
\usepackage{amsmath}
\usepackage{amssymb}
\usepackage{mathtools}
\usepackage{siunitx}
\usepackage{amsfonts}
\usepackage{dsfont}
\usepackage{array}
\usepackage{bm}
\usepackage{mathrsfs}
\usepackage{pifont}
\usepackage{multirow}
\usepackage{upgreek}
\usepackage[dvipsnames]{xcolor}
\usepackage[pdftex,
  pdftitle={Automatic O(a) Improvement for Wilson Twisted Mass Fermions in the Hamiltonian Formulation},
  pdfauthor={Carsten Urbach},
  bookmarks,
  colorlinks,
  linkcolor=myblue,
  citecolor=mymagenta,
  menucolor=black,
  urlcolor=myblue,
  plainpages=false,
  pdfpagelabels,
  hypertexnames=false]{hyperref}
\usepackage{verbatim}
\usepackage{slashed}
\usepackage{cleveref}
\usepackage{ucs}
\usepackage{subfigure}
\usepackage{longtable,booktabs}
\usepackage{braket}

\definecolor{mymagenta}{RGB}{200, 0, 100}
\definecolor{myblue}{RGB}{45, 48, 146}

\graphicspath{{plots/}}

\begin{document}
\title{Automatic $O(a)$ Improvement for Wilson Twisted Mass Fermions
  in the Hamiltonian Formulation}
\author{Carsten Urbach}
\affiliation{Helmholtz-Institut für Strahlen- und Kernphysik, University of Bonn, Nussallee 14-16, 53115 Bonn, Germany}
\affiliation{Bethe Center for Theoretical Physics, University of Bonn, Nussallee 12, 53115 Bonn, Germany}
\date{\today}
\begin{abstract}
  We apply the recently developed Symanzik effective Hamiltonian
  framework to Wilson twisted mass fermions. The effective Hamiltonian
  is constructed solely based on the symmetries of the lattice Hamiltonian
  without recourse to the Symanzik effective action in Euclidean
  space-time. We use it to prove automatic $O(a)$
  improvement at maximal twist, as well as $O(a)$ improvement via
  so-called mass and Wilson average. Moreover, we derive the structure
  of $O(a)$ counterterms pre-requisite for clover on-shell
  improvement of Wilson fermions. We show how to remove on-shell
  redundant contributions using canonical transformations, and find
  that the well known clover term in isotropic Euclidean space time
  splits into two terms with independent coefficients. 
\end{abstract}

\maketitle

\section{Introduction}

The lattice approach provides a highly successful
non-perturbative regularisation for quantum field theories.
This is particularly true for quantum chromodynamics, the theory of
strong interactions, which is non-perturbative at low energies.
One of the main systematic uncertainties to control in lattice quantum
field theories is the extrapolation in the lattice spacing towards the
continuum limit. In particular in lattice quantum chromodynamics
(LQCD), non-perturbative improvement plays, therefore, a crucial role
for the success of the LQCD community in producing results with
sub-percent uncertainties.

Non-perturbative improvement in lattice quantum field theories goes
back to seminal, originally purely perturbative work by
Symanzik~\cite{Symanzik:1981hc,Symanzik:1983dc,Symanzik:1983gh}, and
is formulated as an 
effective field theory in terms of an effective action. This was the
natural framework for lattice field theories, because these are
typically re-formulated in Euclidean space time in order to facilitate
Monte Carlo simulations to solve the path integral. Based on
Symanzik's work, Weisz~\cite{Weisz:1982zw}, Weisz and
Wohlert~\cite{Weisz:1983bn}, Lüscher and Weisz~\cite{Luscher:1984xn}
derived the counter terms for pure gauge theory, followed by
Sheikholeslami and Wohlert~\cite{Sheikholeslami:1985ij}, and Lüscher,
Sint, Sommer, and Weisz~\cite{Luscher:1996sc} for LQCD including
Wilson fermions. 

Symanzik's effective action is derived solely based on dimensional
arguments and the symmetries of the system. Thus, for any given
lattice action a corresponding Symanzik effective action can be
derived in terms of continuum fields. This guiding principle led to
the remarkable work by Frezzotti and Rossi on improvement strategies
using Wilson twisted mass fermions, also in combination with
the so called Osterwalder-Seiler
discretisation~\cite{Frezzotti:2003ni,Frezzotti:2004wz}, which has
been first shown to work in practice in Ref.~\cite{Jansen:2003ir}.

Recently, the Hamiltonian formulation of lattice gauge theories has
attracted more interest again, because tensor network state methods
have been developed significantly, and quantum computers promise to
become available soon. Therefore, the question arises whether a
Symanzik effective Hamiltonian framework can be formulated, for which
we made a proposal in Ref.~\cite{Steffens:2026doc}: again, the effective
Hamiltonian is defined based on dimensional arguments and
symmetries. In Ref.~\cite{Steffens:2026doc} we provided the tools for
how the equivalent of the results by Lüscher and Weisz for pure gauge
theories can be in principle derived directly in the Hamiltonian
formalism based on this framework.

In this paper we go beyond pure gauge theories and apply the framework
to the Hamiltonian of Wilson twisted mass
fermions~\cite{Frezzotti:2000nk} coupled to non-Abelian gauge
fields. In particular, we show that in this 
framework automatic $O(a)$ improvement~\cite{Frezzotti:2003ni} is at
work for Wilson twisted mass fermions at maximal twist in the
Hamiltonian formulation. Moreover, we re-derive the so called 
clover or Sheikholeslami-Wohlert term~\cite{Sheikholeslami:1985ij},
which splits into two terms in 
the effective Hamiltonian. This represents the basis for a future
non-perturbative improvement programme for Wilson clover fermions.

To our knowledge, the first derivation of a classically improved
Hamiltonian including fermions was constructed in
Ref.~\cite{Luo:1993af}, followed by
Refs.~\cite{Mace:2016shq,Spitz:2018eps}. These  
constructions, however, differ significantly from the one presented
here as they represent different ways of \emph{classical}
improvement. Automatic $O(a)$ improvement has been already used in
practice in the Hamiltonian formalism in $d=1$ dimensions in the
Schwinger model in Ref.~\cite{Schwagerl:2025dsj}, but a proof native 
to the Hamiltonian formalism was missing so far. Related work aiming at
spectral reconstruction can be found in
Ref.~\cite{Buividovich:2025hmn}.

\section{Wilson Twisted Mass Hamiltonian}

We start from the continuum Hamiltonian of two flavours of Dirac
fermions coupled to an SU$(N_c)$ gauge field in temporal gauge
\begin{equation}
  \begin{split}
    H_{\mathrm{cont}}
    =&
    \int \mathrm{d}^3x\,
    \hat\Psi^\dagger(\mathbf{x})
    \left(
    -i\boldsymbol{\alpha}\cdot\mathbf D
    +
    M\beta
    \right)
    \hat\Psi(\mathbf{x})\\
    &+\;
    \int \mathrm{d}^3x\,
    \mathrm{tr}\left(\mathbf E^2+\mathbf B^2\right)\,,\\
  \end{split}
\end{equation}
with fermion field operators $\hat\Psi$ obeying standard
anti-commutation relations, quark mass $M$, and the covariant spatial
derivative $\mathbf D=\boldsymbol\nabla-ig\mathbf A$. Physical states
satisfy Gauss' law as a consequence of temporal gauge. We use the
Dirac representation with 
\begin{equation}
  \beta=\gamma^0,
  \qquad
  \alpha_i=\gamma^0\gamma^i\,,
  \qquad
  \gamma_5=i\gamma^0\gamma^1\gamma^2\gamma^3\,.
\end{equation}
In the continuum, the $\omega$-dependent unitary transformation
\begin{equation}
  \begin{split}
    \hat{\Psi}\ & \to\ \exp(i\omega\gamma_5\tau_3/2)\,\hat\Psi\,,\\
    \hat{\Psi}^\dagger\ &\to\ \hat{\Psi}^\dagger\,\exp(-i\omega\gamma_5\tau_3/2)\,,
  \end{split}
\end{equation}
where $\tau_3$ is the Pauli matrix acting in flavour space, is a
symmetry of the massless Hamiltonian, in complete analogy to the
corresponding continuum action~\cite{Frezzotti:2000nk}. Since
$\gamma_5$ anti-commutes with $\beta$ and commutes with $\alpha_i$, it
only leads to a re-parametrisation of the mass term
\begin{equation}
  M\beta\ \to\ M\beta\, \exp(i\omega\gamma_5\tau_3)
  = \beta\left(m_q + i\mu_q\gamma_5\tau_3\right)\,,
\end{equation}
with
\begin{equation}
  m_q = M\cos\omega\,,\quad \mu_q=M\sin\omega\,,
\end{equation}
with $\omega=\pi/2$ corresponding to maximal twist.

For discretising this Hamiltonian, we work on a three dimensional
spatial lattice with sites
\begin{equation}
  \mathbf{n}=(n_1,n_2,n_3)\, ,
  \qquad
  n_i=0,1,\dots,N_i-1\,,
\end{equation}
and total number of sites $N=N_1N_2N_3$. We use periodic boundary
conditions such that, with unit vectors $\mathbf e_i$ in spatial
direction $i\in\{1,2,3\}$,
\begin{equation}
  \mathbf n+\mathbf e_i
  \equiv
  \mathbf n+\mathbf e_i \pmod{N_i}\,.
\end{equation}
Physical coordinates $\mathbf x_{\mathbf n}=a\mathbf n$ are obtained
by multiplying with the lattice spacing $a$. The corresponding lattice
Hamiltonian for the fermions, the so-called Wilson twisted mass
Hamiltonian, reads
\begin{equation}
  \label{eq:HWTMgauge}
  \begin{aligned}
    H_{\text{WTM}}
    &=
    \sum_{\mathbf n}
    \Psi_{\mathbf n}^\dagger
    \beta\left(m_q + i\mu_q\gamma_5\tau_3\right)
    \Psi_{\mathbf n}
    \\[1ex]
    -&
    \nu\frac{i}{2a}
    \sum_{\mathbf n,i}
    \Psi_{\mathbf n}^\dagger
    \alpha_i
    \left(U_{\mathbf n, i}
    \Psi_{\mathbf n+\mathbf e_i}
    -
    U_{\mathbf n-\mathbf e_i, i}^\dagger\Psi_{\mathbf n-\mathbf e_i}
    \right)
    \\[1ex]
    +&
    \nu\frac{r}{2a}
    \sum_{\mathbf n,i}
    \Psi_{\mathbf n}^\dagger
    \beta
    \left(
    2\Psi_{\mathbf n}
    -
    U_{\mathbf n, i}\Psi_{\mathbf n+\mathbf e_i}
    -
    U_{\mathbf n-\mathbf e_i, i}^\dagger\Psi_{\mathbf n-\mathbf e_i}
    \right)\,.
  \end{aligned}
\end{equation}
Here, $\Psi_{\mathbf n}=(\psi_{\mathbf{n},1},\ \psi_{\mathbf{n},2})^t$
is a two flavour fermion operator with
\begin{equation}
  \{\psi_{\mathbf n\alpha fc},\psi_{\mathbf m\alpha'f'c'}^\dagger\}
  =
  \delta_{\mathbf n\mathbf m}\delta_{\alpha\alpha'}\delta_{ff'}\delta_{cc'}\,,
\end{equation}
at sites $\mathbf{n},\mathbf{m}$ with $f=1,2$ the flavour index,
$\alpha=1,\dots,4$ the spinor index and $c=1,\dots,N_c$ the colour
index; $\nu$ is related to the fermion velocity, which we cannot
absorb in the field normalisation, because it is fixed by the
anti-commutation relations. $r$ is the Wilson parameter.
$\Psi$ is related to the continuum-normalised field via 
\begin{equation}
  \Psi_{\mathbf{n}} = \sqrt{a^3}\, \hat\Psi(\mathbf{x}_\mathbf{n})\,.
\end{equation}
$m_q$ is called the bare Wilson quark mass, whereas $\mu_q$ is the
bare twisted quark mass. The first line of \cref{eq:HWTMgauge}
contains the Wilson and twisted mass terms, the second the kinetic
term, and the third the Wilson term removing the fermion doublers.

$U_{\mathbf n,i}\in\mathrm{SU}(N_c)$ denotes the gauge link operator on
the link from $\mathbf n$ to $\mathbf n+\mathbf e_i$, $i=1,2,3$. The
plaquette operator is defined from these as
\begin{equation}
  U_{\mathbf n, ij} = U^{~}_{\mathbf n,i}\, U^{~}_{\mathbf n +\mathbf e_i,j}\,
  U^\dagger_{\mathbf n + \mathbf e_j,i}\, U^\dagger_{\mathbf n,j}\,.
\end{equation}
On each link we introduce left and right generators
$L^a_{\mathbf n,i}$ and $R^a_{\mathbf n,i}$, $a=1,\dots,N_c^2-1$,
with
\begin{equation}
  \begin{split}
    [L^a_{\mathbf n,i},U_{\mathbf m,j}] &= -\delta_{\mathbf n\mathbf m}\delta_{ij}\,T^aU_{\mathbf n,i}\,,\\
    [R^a_{\mathbf n,i},U_{\mathbf m,j}] &= \delta_{\mathbf n\mathbf m}\delta_{ij}\,U_{\mathbf n,i}T^a\,,\\
    [L^a_{\mathbf n,i},L^b_{\mathbf m,j}] &= i\delta_{\mathbf n\mathbf m}\delta_{ij}f^{abc}L^c_{\mathbf n,i}\,,\\
    [R^a_{\mathbf n,i},R^b_{\mathbf m,j}] &= i\delta_{\mathbf n\mathbf m}\delta_{ij}f^{abc}R^c_{\mathbf n,i}\,,
  \end{split}
\end{equation}
and $[L^a,R^b]=0$, where $T^a$ are the Hermitian generators of
SU$(N_c)$ normalised as $\mathrm{tr}\,T^aT^b=\delta^{ab}/2$. On each
link $L^a_{\mathbf n,i}L^a_{\mathbf n,i}=R^a_{\mathbf n,i}R^a_{\mathbf n,i}$.
The standard Kogut-Susskind Hamiltonian~\cite{Kogut:1974ag} for the
pure gauge part reads
\begin{equation}
  H_g = c_0\left(\frac{g^2}{2a}\sum_{\mathbf n, i} L^a_{\mathbf n,i}L^a_{\mathbf n,i} +
  \frac{2}{g^2a}\sum_{\mathbf n, i<j}\mathrm{Re\, tr}\left(1-U_{\mathbf n, ij}\right)\right)\,,
\end{equation}
and the full Hamiltonian is $H=H_{\text{WTM}}+H_g$. To leading order
in $a$ we identify $L^a_{\mathbf n,i}\simeq
-\frac{a^2}{g}E^a_i(\mathbf x_{\mathbf n})$ and $R^a_{\mathbf
  n,i}\simeq +\frac{a^2}{g}E^a_i(\mathbf x_{\mathbf n})$. 
Physical states are annihilated by the Gauss law operator
\begin{equation}
  G^a_{\mathbf n}
  =
  \sum_i\left(L^a_{\mathbf n,i}+R^a_{\mathbf n-\mathbf e_i,i}\right)
  + \Psi^\dagger_{\mathbf n}T^a\Psi_{\mathbf n}\,,
\end{equation}
where the sign of the charge term $\Psi^\dagger T^a\Psi$ follows the
conventions for $L$ and $R$ above. 

As discussed in Ref.~\cite{Steffens:2026doc}, the speed of light $c_0$
must be included in $H_g$ due to the breaking of Lorentz
invariance. The equivalent parameter in the fermionic Hamiltonian
\cref{eq:HWTMgauge} is $\nu$, and represents a $O(a^0)$ correction to the fermion
velocity, with $\nu=1$ at tree level. Only the ratio $c_0/\nu$ is
physical and needs to be tuned. Note that for the Wilson term 
$\nu$ could be absorbed in $r$.

\subsection{Charge Conjugation, Parity, Time Reversal}

Let us define charge conjugation, parity and time reversal for the
Wilson twisted mass Hamiltonian. We choose a basis of generators with
$(T^a)^\star = s_a T^a$, $s_a = \pm 1$ (e.g.\ the Gell-Mann basis, where
$s_a=+1$ for the real symmetric and $s_a=-1$ for the imaginary
anti-symmetric generators). Charge conjugation is unitary and acts as a
particle-hole transformation. With $V_c=\gamma^2$ in the Dirac basis,
ordinary charge conjugation is acting as
\begin{equation}
  \mathcal C:\quad
  \begin{cases}
    \Psi_{\mathbf n}\ &\to\ V_c\,\big(\Psi^\dagger_{\mathbf n}\big)^t\,,\\
    \Psi^\dagger_{\mathbf n}\ &\to\ \Psi^t_{\mathbf n}\,V_c^\dagger\,,\\
    U_{\mathbf n,i}\ &\to\ U_{\mathbf n,i}^\star\,,\\
    L_{\mathbf n,i}^a\ &\to\ -s_a L_{\mathbf n,i}^a\,,\\
    R_{\mathbf n,i}^a\ &\to\ -s_a R_{\mathbf n,i}^a\,,
  \end{cases}
\end{equation}
where the transposition acts on spinor and flavour indices and the
colour index is carried over $\mathcal C$ is a symmetry independently
of $m_q$ 
and $\mu_q$. Ordinary parity, on the other hand, needs to be replaced
by twisted parity
\begin{equation}
  \mathcal P_\mathrm{tm}:\quad
  \begin{cases}
    \Psi_{\mathbf n}\ &\to\ V_p\,\Psi_{-\mathbf n}\,,\\
    \Psi^\dagger_{\mathbf n}\ &\to\ \Psi^\dagger_{-\mathbf n}\,V_p^\dagger\,,\\
    U_{\mathbf n,i}\ &\to\ U^\dagger_{-\mathbf{n}-\mathbf e_i, i}\,,\\
    L_{\mathbf n, i}^a\ &\to\ R_{-\mathbf n-\mathbf e_i, i}^a\,,\\
    R_{\mathbf n, i}^a\ &\to\ L_{-\mathbf n-\mathbf e_i, i}^a\,,
  \end{cases}
  \qquad
  V_p = \beta\tau_1 \ \text{ or }\ \beta\tau_2\,.
\end{equation}
Ordinary parity, $V_p=\beta$, is only a symmetry at $\mu_q=0$.
Time reversal $\mathcal T$ must be anti-unitary in the Hamiltonian
formulation, $\mathcal T\,z\,\mathcal T^{-1}=z^\star$ for complex
numbers $z$. It also needs to be combined with a flavour rotation to
remain a symmetry of the twisted Hamiltonian. In the Dirac basis we
write
\begin{equation}
  \mathcal T_\mathrm{tm} = V_t\, \mathcal K\,,
\end{equation}
with $\mathcal K$ the complex conjugation of the coefficients of a
state, and
$V_t = \gamma^1\gamma^3\tau_{1}$ or $\gamma^1\gamma^3\tau_2$ acting on
spinor and flavour indices. The action of $\mathcal T_\mathrm{tm}$
on the operators reads
\begin{equation}
  \mathcal{T}_\mathrm{tm}:\quad
  \begin{cases}
    \Psi_{\mathbf n}\ &\to\ V_t\,\Psi_{\mathbf n}\,,\\
    \Psi^\dagger_{\mathbf n}\ &\to\ \Psi^\dagger_{\mathbf n}V_t^\dagger\,,\\
    U_{\mathbf n,i} \ &\to\ U_{\mathbf n,i}\,,\\
    L_{\mathbf n,i}^a\ &\to\ s_a L_{\mathbf n,i}^a\,,\\
    R_{\mathbf n,i}^a\ &\to\ s_a R_{\mathbf n,i}^a\,,
  \end{cases}
\end{equation}
where the link transformation holds entry-wise. Like for parity,
ordinary time reversal, $V_t=\gamma^1\gamma^3$, is only a symmetry at
$\mu_q=0$. In all three cases the Gauss law operator transforms
covariantly, $G^a_{\mathbf n}\to -s_ag^a_{\mathbf n}$,
$G^a_{\mathbf n}\to G^a_{-\mathbf n}$ and $G^a_{\mathbf n}\to s_aG^a_{\mathbf n}$
for $\mathcal C$, $\mathcal P_\mathrm{tm}$ and $\mathcal T_\mathrm{tm}$,
respectively, such that the physical subspace is preserved. The
flavour matrices in $V_p$ and $V_t$ are fixed only up to the exact
$U(1)_3$ symmetry, i.e.\ $\tau_1\to\tau_1e^{i\theta\tau_3}$.

\section{$O(a)$ Improvement}

\subsection{Effective Hamiltonian with Wilson-type fermions}

As discussed in Ref.~\cite{Steffens:2026doc}, the Symanzik effective
Hamiltonian is of the form
\begin{equation}
  \label{eq:Heff}
  \begin{split}
    H_{\rm eff}&=\sum_{k\ge -1}a^k H_k\,,\\
    H_k&=\sum_i c_{k,i}(r)\,\int \mathrm{d}^dx\,\mathcal
    O_{k,i}(\mathbf x)\,,\\
  \end{split}
\end{equation}
where $a\equiv a_s$ is the spatial lattice spacing. We will assume
$d=3$ in the following, but the arguments go through with $d=1$ as
well. The densities
$\mathcal O_{k,i}$ are Hermitian, local, gauge invariant and of mass
dimension $d+1+k$. They are built from the continuum fields, spatial
covariant derivatives, $\mathbf E$ and $\mathbf B$, and they include
monomials in the bare masses as spurions. The coefficients
$c_{k,i}$ are real due to the hermiticity of the densities, depend on
the Wilson parameter $r$, on $g^2$, and possibly 
on $\log a$. The operators are restricted by dimensional analysis and
by the symmetries of the lattice Hamiltonian. We choose the basis
such that each $\mathcal O_{k,i}$ has a definite transformation
behaviour under the transformations defined below. This is always
possible since all of them square to the identity, and it is
compatible with the redundancies, i.e.\ total derivatives, Gauss
law and unitary transformations $\exp(ia^kX)$.

With Wilson-type fermions chiral symmetry is explicitly broken, and the
untwisted quark mass renormalises additively. This appears as the
$k=-1$ power divergent term, $\frac1a c_{-1}(r)\,\bar\psi\psi$, which
is of course absent in pure gauge theory.
This additive renormalisation is
expressed via the critical mass $m_\mathrm{crit}(r)$ and the subtracted
bare mass
\begin{equation}
  \label{eq:mhatdef}
  \hat m(m_q, r) = m_q - m_\mathrm{crit}(r)\,,
\end{equation}
where $m_\mathrm{crit}$ can for instance be fixed by the
renormalisation condition 
\begin{equation}
  \label{eq:mCrit}
  \langle 0|A_0^{\pi^+}|\pi^+(0)\rangle = 0\,,
\end{equation}
with
\begin{equation}
  \label{eq:A0}
  A_0^{\pi^+} = \sum_x \bar d(x)\,i\,\gamma_0\gamma_5\, u(x)
\end{equation}
the time component of the axial-vector current.

For the pure gauge part the result of Ref.~\cite{Steffens:2026doc}
carries over: CPT forbids all dimension-5 pure gauge operators, so
gauge field corrections start at $O(a^2)$. With fermions there are
$O(a)$ terms, but every one of them contains fermion fields or a
factor of the quark mass. For instance, $\hat m\,\mathrm{tr}\,\mathbf E^2$
and $\hat m\,\mathrm{tr}\,\mathbf B^2$ are allowed, while
$\mu_q\,\mathrm{tr}\,\mathbf E^2$ and $\mu_q\,\mathrm{tr}\,\mathbf B^2$
are forbidden by flavour exchange, see below. In turn, the additional
bare parameters $\{r,m_q,\mu_q\}$ allow us to relate the coefficients
$c_{k,i}$ at different parameter values. The arguments follow closely
the ones by Frezzotti and Rossi in 4d Euclidean space
time~\cite{Frezzotti:2003ni}.

\subsection{Spurionic transformations}

In addition to $\mathcal C$, $\mathcal P_\mathrm{tm}$ and
$\mathcal T_\mathrm{tm}$ defined above we consider the following
transformations, which do not affect the gauge fields:
\begin{equation}
  \label{eq:spurions}
  \begin{aligned}
    \mathcal R_5:&\quad \Psi_\mathbf{n}\to\gamma_5\Psi_\mathbf{n}\,,\\
    \mathcal R_5':&\quad \Psi_\mathbf{n}\to\gamma_5\tau_1\Psi_\mathbf{n}\,,
  \end{aligned}
\end{equation}
with $\Psi^\dagger_\mathbf{n}$ transforming accordingly.
It follows from $\gamma_5\beta\gamma_5=-\beta$,
$\gamma_5\alpha_i\gamma_5=\alpha_i$ and $\tau_1\tau_3\tau_1=-\tau_3$.
These transformations change bare parameters and are therefore
spurionic symmetries: e.g.\
$\mathcal R_5\times[r\to-r]\times[m_q\to-m_q]\times[\mu_q\to-\mu_q]$
leaves the lattice Hamiltonian invariant, and, likewise,
$\mathcal
R_5^\prime\times[r\to-r]\times[m_q\to-m_q]\times[\mu_q\to\mu_q]$.
Let us also introduce the combination
\begin{equation}
  \label{eq:S}
  \mathcal{S} = \mathcal{P}_\mathrm{tm} \mathcal{R}_5^\prime\,,
\end{equation}
which will be important for the proof of $O(a)$ improvement at maximal
twist. Finally, we use the CPT transformation
$\Theta\equiv\mathcal C\,\mathcal P_\mathrm{tm}\,\mathcal T_\mathrm{tm}$.
It is an exact, anti-unitary, flavour diagonal symmetry with spinor
matrix proportional to $\gamma_5$. The combination of $\mathcal R_5$
and $\mathcal R_5^\prime$ will be denoted as $\tau_1\equiv \mathcal
R_5\mathcal R_5^\prime$. 

We make the important assumption that the identification of the low energy sector of the
lattice Hilbert space with the continuum Fock space can be chosen
covariant under these transformations. Then the relations
\cref{eq:spurions} also hold for $H_\mathrm{eff}$.

\subsection{Chirality and CPT}

For a fermion bilinear $\bar\psi\Gamma\psi$, let $q$ be the
number of $\gamma^\mu$ in $\Gamma$, with $\gamma_5$ counting as even
and $\bar\psi = \psi^\dagger\gamma_0$.
Its $\mathcal R_5$-chirality is
\begin{equation}
  \chi(\bar\psi\Gamma\psi) = (-1)^{q+1}\,.
\end{equation}
Gauge fields and covariant derivatives have $\chi=+1$. For an operator
containing spurions, $\chi$ denotes the total sign under $\mathcal R_5$
including the sign flips of the parameters. Thus $m_q$ and $\mu_q$,
and also $r$, carry $\chi=-1$, while $a$ and $g^2$ carry $\chi=+1$.

The CPT sign of a Hermitian operator $O$ with real coefficient is
\begin{equation}
  \theta(O)=(-1)^{\mathrm{rank}(O)}\,,
\end{equation}
where $\mathrm{rank}$ counts every Lorentz index carried by
$\gamma$-matrices, covariant derivatives and field strength tensors,
including indices contracted with the time-like unit vector $n_\mu$.
$F_{\mu\nu}$ counts two indices (including any $n_\mu$), $D$ and
$\gamma^\mu$ one each, and $\epsilon$ none. For fermion
bilinears this follows from the action of $\Theta$ on
$\bar\psi\Gamma D^kF^f\psi$, which gives $(-1)^{q+k}$. Denoting the
mass dimension in $d=3$ by $d_m$, our finding for the building blocks
is compiled in \cref{tab:chitheta}. Note that in that table we refer
to Hermitian bilinears $\bar\psi\Gamma\psi$.

\begin{table}[h]
  \centering
  \begin{tabular*}{\linewidth}{@{\extracolsep{\fill}}lcccc}
    \hline
    & $d_m$ & $\chi$ & $\theta$ & $\chi\,(-1)^{d_m}$ vs.\ $\theta$\\
    \hline\hline
    $\psi^\dagger\beta\Gamma^{~}\psi$ & $3$ & $(-1)^{q+1}$ & $(-1)^{q}$ & equal\\
    $D_i$ & $1$ & $+1$ & $-1$ & equal\\
    $E_i$, $B_i$ & $2$ & $+1$ & $+1$ & equal\\
    $m_q$, $\hat m$, $\mu_q$ & $1$ & $-1$ & $+1$ & equal\\
    $g^2$ & $0$ & $+1$ & $+1$ & equal\\
    \hline
    $a$ & $-1$ & $+1$ & $+1$ & opposite\\
    $r$ & $0$ & $-1$ & $+1$ & opposite\\
    \hline
  \end{tabular*}
  \caption{Mass dimension, $\mathcal R_5$-chirality and CPT sign of
    the building blocks of $H_\mathrm{eff}$ in $d=3$. For $\hat m$
    the entry anticipates \cref{eq:mhat}.}
  \label{tab:chitheta}
\end{table}

\noindent Since all quantities are multiplicative, every monomial $O$
built from the field-theoretic building blocks and the masses satisfies
\begin{equation}
  \label{eq:chitheta}
  \chi(O)\,(-1)^{d_m(O)}=\theta(O)\,,
\end{equation}
with the only exceptions being the lattice parameters $a$ and $r$.

\subsection{Wilson average}

Consider now a term $a^k c_k(r)\,O$ in $H_\mathrm{eff}$. Here $O$ contains
all fields and mass monomials in $m_q$ and $\mu_q$.
Since $\Theta$ is an exact symmetry and $c_k$ is real,
$\theta(O)=+1$. With $d_m(O)=d+1+k$, \cref{eq:chitheta} yields
\begin{equation}
  \label{eq:chik}
  \chi(O)=(-1)^{d+1+k}=(-1)^k\qquad (d=3)\,.
\end{equation}
Write $\chi(O)=\chi_f(O)\,\chi_s(O)$. Here $\chi_f$ is the sign of the
field part under $\mathcal R_5$, and $\chi_s=(-1)^{n_m}$ with $n_m$
the number of factors $m_q$ and $\mu_q$ in the mass monomial, so that
$O(-m_q,-\mu_q)=\chi_s(O)\,O(m_q,\mu_q)$. Applying $\mathcal R_5$ to
$H_\mathrm{eff}$ and using \cref{eq:spurions} gives
\begin{equation}
  \begin{split}
    \sum_{k,i}&a^kc_{k,i}(r)\,\chi_f(O_{k,i})\,O_{k,i}(m_q,\mu_q)\\
    &=\sum_{k,i}a^kc_{k,i}(-r)\,\chi_s(O_{k,i})\,O_{k,i}(m_q,\mu_q)\,.\\
  \end{split}
\end{equation}
Comparing coefficients of linearly independent operators and powers of
$a$, which $\mathcal R_5$ does not act on, we obtain with \cref{eq:chik}
\begin{equation}
  \label{eq:WA}
  c_{k,i}(-r)=(-1)^k\,c_{k,i}(r)\qquad\forall\, k\ge-1\,,
\end{equation}
under the assumptions listed above. The same argument applies to the
expansion of 
a local lattice operator with definite $\mathcal R_5$ and $\Theta$
transformation behaviour,
\begin{equation}
  \begin{split}
    O_\mathrm{lat}=&Z_O\Big(O_0+\sum_{k\ge1}a^k\sum_i d_{k,i}(r)\,O_{k,i}\Big)\,,\\
    &d_{k,i}(-r)=(-1)^k\,d_{k,i}(r)\,.\\
  \end{split}
\end{equation}
For $k=-1$, \cref{eq:WA} implies $m_\mathrm{crit}(-r)=-m_\mathrm{crit}(r)$
and therefore
\begin{equation}
  \label{eq:mhat}
  \hat m(-r,-m_q)=-\hat m(r,m_q)\,.
\end{equation}
Hence $\hat m$ is a consistent spurion with $\chi=-1$, and from now on
we organise the expansion in $\hat m$ and $\mu_q$. For a physical
quantity $P(r,\hat m,\mu_q)=P_0(\hat m,\mu_q)+\sum_{k\ge1}a^kp_k(r,\hat m,\mu_q)$,
\cref{eq:WA} implies $p_k(-r)=(-1)^kp_k(r)$. This is the Wilson average
of Ref.~\cite{Frezzotti:2003ni}:
\begin{equation}
  \tfrac12\big[P(r,\hat m,\mu_q)+P(-r,\hat m,\mu_q)\big]=P_0(\hat m,\mu_q)+O(a^2)\,,
\end{equation}
free of all odd powers of $a$. Note that the average has to be taken
at fixed $\hat m$, i.e.\ with bare masses $m_q=\hat m\pm m_\mathrm{crit}(r)$.

\subsection{Mass average}

Applying $\mathcal R_5'$ in the same way, with $\chi_f'$ the field
sign under $\mathcal R_5'$ and using \cref{eq:WA,eq:mhat}, gives for
every operator with $c_{k,i}\neq0$
\begin{equation}
  \label{eq:MAop}
  \begin{split}
    \mathcal R_5'\,O_{k,i}(\hat m,\mu_q)\,\mathcal R_5'^{-1}
    &=\chi_f'(O_{k,i})\,O_{k,i}(\hat m,\mu_q)\\
    &=(-1)^k\,O_{k,i}(-\hat m,\mu_q)\,.\\
  \end{split}
\end{equation}
Operators at the same order $k$ may have different $\chi_f'$. Then the
compensating factor is provided by the parity of their $\hat m$
dependence. The right-hand side, however, is universal. Summing over
all terms, we find
\begin{equation}
  \label{eq:MA}
  H_\mathrm{eff}(a,-\hat m,\mu_q)=\mathcal R_5'\,H_\mathrm{eff}(-a,\hat m,\mu_q)\,\mathcal R_5'^{-1}\,,
\end{equation}
where $H_\mathrm{eff}(-a)$ denotes the formal expansion with
$a^k\to(-a)^k$. The Hamiltonians at $-\hat m$ and at $\hat m$ with the
sign of $a$ flipped are therefore unitarily equivalent. In particular,
$E_n(a,-\hat m)=E_n(-a,\hat m)$ for the energy levels, where
$\mathcal R_5'$ maps the eigenstates onto each other. $\mathcal R_5'$
exchanges flavours, e.g.\ $\pi^+\leftrightarrow\pi^-$, but these
states are degenerate due to $\mathcal C$. Alternatively one can use
the flavour diagonal combination $\mathcal S$ \cref{eq:S}, under which
$H(r,\hat m,\mu_q)\to H(-r,-\hat m,\mu_q)$ holds as well. Consequently,
the mass average
\begin{equation}
  \tfrac12\big[P(r,\hat m,\mu_q)+P(r,-\hat m,\mu_q)\big]=P_0(\hat m,\mu_q)+O(a^2)
\end{equation}
is free of odd powers of $a$ for energies and for matrix elements of
$\mathcal S$-even operators. For $\mathcal S$-odd operators the mass
difference is improved instead.

\subsection{$O(a)$ improvement at maximal twist}

At maximal twist, $\hat m=0$ and all quark mass dependence is in
$\mu_q$. For $\mathcal R_5^\prime$-even observables we can infer from
mass average that
\begin{equation}
  p_k(-\hat m) = (-1)^k p_k(\hat m)\,.
\end{equation}
Since for $\mu_q\neq0$ these $p_k$ are analytic in $\hat m$ and in
particular at $\hat m = 0$, all $k$-odd coefficients $p_k$ must vanish
at $\hat m =0$, i.e. at maximal twist. The operators in $H_1$ do not
vanish by themselves, however, \cref{eq:MAop} becomes
\begin{equation}
  \label{eq:SHk}
  \mathcal S\,H_k\,\mathcal S^{-1}=(-1)^k\,H_k\qquad(\hat m=0)\,,
\end{equation}
i.e.\ $H_0$ is invariant under $\mathcal S$, while all terms with odd
$k$, in particular $H_1$, are $\mathcal S$-odd. For eigenstates of
$\mathcal S$, every contribution of order $a^k$ to an energy level
has $\mathcal S$-parity $(-1)^k$ in perturbation theory in $a$.
Hence, we have for energy differences of non-degenerate states
\begin{equation}
  \label{eq:En}
  E_n=E_n^{(0)}+O(a^2)\,,
\end{equation}
and the same holds for matrix elements of $\mathcal S$-even
operators. Matrix elements that vanish in the continuum by
$\mathcal S$ contain only odd powers of $a$. An example is the
temporal axial current between vacuum and pion \cref{eq:mCrit}, which vanishes in
the continuum at maximal twist and is used to tune
$\hat m=0$~\cite{Frezzotti:2003ni}. Thus $O(a)$ improvement is
automatic at maximal twist, without any improvement coefficient in
the Hamiltonian or in the operators. This remains true if $\hat m$ is
tuned to zero only up to $O(a)$ corrections, since $\hat m\,\bar\psi\psi$
is $\mathcal S$-odd.
Such a residual $\hat m=O(a)$ may, however, enhance $O(a^2)$ effects
at small $\mu_q$, see
Refs.~\cite{Aoki:2004ta,Sharpe:2004ny,Frezzotti:2005gi}. A safe choice
is, for instance, to ensure $\mu_q/\hat m$ is large enough for any
residual $\hat m$, as shown to work by the Extended Twisted Mass
Collaboration~\cite{ETM:2008zte}. 

Let us finish this section with a remark on degenerate energy levels
and their improvement. 
At $\hat m=0$, \cref{eq:SHk} in fact only implies $\mathrm{spec}
\{H_\mathrm{eff}(a)\} = \mathrm{spec}\{ H_\mathrm{eff}(-a)\}$. Thus,
there might be accidental cases where \cref{eq:En} does not hold in
case of degenerate energy levels.

\section{$H_1$ Operator Classification}

\begin{table*}[t]
  \centering
  \begin{tabular}{clcc}
    \hline
    & Hamiltonian density & origin & $\mathcal{S}$ at $\hat m=0$ \\
    \hline\hline
    $O_1$ & $\psi^\dagger\beta D^2\psi$                          & Wilson-type, spatial   & odd \\
    $O_2$ & $\psi^\dagger\beta\,\Sigma_i B_i\,\psi$              & Pauli, magnetic        & odd \\
    $O_3$ & $\psi^\dagger i\beta\,\alpha_i E_i\,\psi$            & Pauli, electric        & odd \\
    $O_4$ & $\hat m\,\psi^\dagger\alpha_i(-iD_i)\psi$            & mass $\times$ kinetic  & --  \\
    $O_5$ & $\hat m\,\mathrm{tr}(E_iE_i)$                        & mass $\times$ electric & --  \\
    $O_6$ & $\hat m\,\mathrm{tr}(B_iB_i)$                        & mass $\times$ magnetic & --  \\
    $O_7$ & $\hat m^2\,\psi^\dagger\beta\psi$                    & $b_m$-type             & --  \\
    $O_8$ & $\mu_q^2\,\psi^\dagger\beta\psi$                     & $b_\mu$-type           & odd \\
    $O_9$ & $\hat m\mu_q\,\psi^\dagger i\beta\gamma_5\tau_3\psi$ & twist-angle shift      & --  \\
    \hline
  \end{tabular}
  \caption{Allowed dimension-5 candidate densities for $H_1$ for $N_f=2$ Wilson twisted mass
    fermions (twisted basis) at generic $(\hat m,\mu_q)$ in Hamiltonian form, with
    $D^2=D_iD_i$, $E_i=E_i^aT^a$, $B_i=B_i^aT^a$, $\Sigma_i=\tfrac12\epsilon_{ijk}\sigma_{jk}$
    and colour indices suppressed in the Pauli terms. All
    coefficients are odd in the Wilson parameter $r$, and the last column gives the
    parity under $\mathcal{S}=\mathcal{P}_{tm}R_5'$ at maximal twist $\hat m=0$
    (``--'': term vanishes at $\hat m=0$). }
  \label{tab:H1}
\end{table*}

\begin{table*}[t]
  \centering
  \begin{tabular}{lcc}
    \hline
    Hamiltonian density & \qquad~\quad~\quad & forbidden by \\
    \hline\hline
    $\mu_q\,\mathrm{tr}(E_iE_i)$, $\mu_q\,\mathrm{tr}(B_iB_i)$         &  & $\tau_1$ \\
    $\hat m\,\mathrm{tr}(E_iB_i)$                                      &  & $\mathcal{P}_\mathrm{tm}$, $\mathcal{T}_\mathrm{tm}$ \\
    $\mu_q\,\mathrm{tr}(E_iB_i)$                                       &  & $\mathcal{P}_\mathrm{tm}$, $\mathcal{T}_\mathrm{tm}$, $\tau_1$ \\
    $\psi^\dagger D^2\psi$                                             &  & $\mathcal{C}$, $\Theta$ \\
    $\psi^\dagger i\beta\gamma_5\tau_3 D^2\psi$                        &  & $\tau_1$ \\
    $\mu_q\,\psi^\dagger\Sigma_i\tau_3(-iD_i)\psi$                     &  & $\mathcal{C}$, $\mathcal{T}_\mathrm{tm}$ \\
    all other one-derivative bilinears see \cref{app:onederiv}         &  & $\mathcal{P}_\mathrm{tm}$, $\mathcal{C}$, $\mathcal{T}_\mathrm{tm}$, $\Theta$ or $\tau_1$ \\
    $\psi^\dagger\alpha_iE_i\psi$, $\psi^\dagger\alpha_iB_i\psi$       &  & $\Theta$ \\
    $\psi^\dagger\Sigma_iE_i\psi$, $\psi^\dagger\Sigma_iB_i\psi$       &  & $\Theta$ \\
    $\psi^\dagger\beta\,\Sigma_iE_i\,\psi$, $\psi^\dagger
    i\beta\,\alpha_iB_i\,\psi$ & & $\mathcal{P}_{tm}$ \\
    $\tau_3$ variants of rows 2 and 3                                  &  & $\tau_1$ \\
    pure-gauge dimension-5 operators                                   &  & $\Theta$ \\
    \hline
  \end{tabular}
  \caption{Like \cref{tab:H1}, but candidates excluded by the
    exact symmetries $\mathcal{P}_{tm}$, $\mathcal{C}$, $\mathcal{T}_\mathrm{tm}$, $\Theta$ (CPT)
    or by the flavour-exchange $\tau_1$ ($\mu_q\to-\mu_q$).}
  \label{tab:H1excl}
\end{table*}
 
The general argument of the previous section establishes automatic
$O(a)$ improvement at maximal twist without listing the operators in
$H_1$. It is nevertheless instructive to classify all operators which
could contribute to $H_1$ for dimensional reasons, and to discuss
explicitly which of them are absent and why they do not contribute at
maximal twist. This is the Hamiltonian counterpart of the proof of
automatic $O(a)$ improvement by Shindler in
Ref.~\cite{Shindler:2007vp}.

The possible contributions to $H_1$, i.e.\ densities of mass dimension
five, are compiled in \cref{tab:H1,tab:H1excl}. 
The former contains the nine
operators $O_1,\dots,O_9$ allowed at generic twist angle, including
the case $\mu_q=0$. The latter lists dimension five operators
excluded by the exact symmetries $\mathcal C$, $\mathcal P_\mathrm{tm}$,
$\mathcal T_\mathrm{tm}$, $\Theta$, or by the flavour exchange spurion
$\tau_1$. For mass-weighted one-derivative bilinears see
\cref{app:onederiv}.

\subsection{On-shell Redundant Operators}

Not all of $O_1,\dots,O_9$ are independent on-shell. A unitary
transformation $e^{ia\epsilon\Phi}$ with a Hermitian, gauge invariant
operator $\Phi$ with mass dimension $1$, changes
\begin{equation}
  \label{eq:redundancy}
  H_1\ \to\ H_1-\epsilon\, i[H_0,\Phi]
\end{equation}
without affecting the spectrum~\cite{Steffens:2026doc}. For $e^{ia\epsilon\Phi}$ to preserve
the exact symmetries, $\Phi$ must be even under $\mathcal C$ and
$\mathcal P_\mathrm{tm}$. It must be odd under the anti-unitary
$\mathcal T_\mathrm{tm}$ and $\Theta$, since
$\mathcal A e^{ia\epsilon\Phi}\mathcal A^{-1}=e^{-ia\epsilon\mathcal A\Phi\mathcal A^{-1}}$
for an anti-unitary $\mathcal A$. In addition, $\Phi$ must be invariant
under $U(1)_3$ and flavour exchange (i.e. $\tau_1$).
We find two such generators, the first of which is
\begin{equation}
  \Phi_1=\int\mathrm{d}^3x\,\psi^\dagger K_1\psi\,,
  \qquad K_1=\beta\alpha_iD_i\,,
\end{equation}
where $K_1$ is Hermitian since $D_i$ is anti-Hermitian. Using the
canonical commutation relations, $\dot A_i=E_i$ and the identities
$\alpha_i\alpha_j=\delta_{ij}+i\epsilon_{ijk}\Sigma_k$,
$[D_i,D_j]=-igF_{ij}$, $\{\beta,\alpha_i\}=\{\beta,\gamma_5\}=0$, one
finds
\begin{equation}
  \label{eq:Phi1}
  i[H_0,\Phi_1]=-2\,O_1-2g\,O_2-g\,O_3-2\,O_4\,.
\end{equation}
The four terms come from the kinetic term of $H_0$ ($O_1$, $O_2$), the
time dependence of the gauge field in $D_i$ ($O_3$), and the Wilson
mass term ($O_4$). The twisted mass contributions cancel. Hence
$\Phi_1$ allows us to remove $O_1$ in favour of $O_2$, $O_3$ and $O_4$.
It generates the canonical transformation
$\psi\to\psi-ia\epsilon\,\gamma^iD_i\psi$, the Hamiltonian counterpart
of the Lagrangian field redefinition $\psi\to\psi+\epsilon\gamma_iD_i\psi$
used to remove $\bar\psi D^2\psi$ on-shell. Accordingly, the expansion
of every lattice operator has to be rotated as well,
\begin{equation}
  O\ \to\ O+i\epsilon a\,[\Phi_1,O]\,.
\end{equation}

The second generator is $\mu_q$ times the non-singlet axial charge,
\begin{equation}
  \Phi_2=\mu_q\int\mathrm{d}^3x\,\psi^\dagger\gamma_5\tau_3\psi=\mu_q\,Q_5^{3}\,.
\end{equation}
It generates an infinitesimal rotation of the twist angle, which is
free of anomalies since $\tau_3$ is traceless. Only the mass terms of
$H_0$ contribute, and
\begin{equation}
  \label{eq:Phi2}
  i[H_0,\Phi_2]=2\,O_9-2\,O_8\,.
\end{equation}
Therefore, $O_8$ or $O_9$ can be removed in favour of the other. This
corresponds to an $O(a\mu_q)$ redefinition of the twist angle.

\subsection{Maximal twist}

At maximal twist, $\hat m=0$, the operators $O_4,\dots,O_7$ and $O_9$
vanish. $O_1$ and $O_8$ can be removed with $\Phi_1$ and $\Phi_2$,
respectively. This leaves
\begin{equation}
  H_1=c_\mathrm{SW}^m\,O_2+c_\mathrm{SW}^e\,O_3\qquad(\hat m=0)\,.
\end{equation}
These terms are present with non-vanishing coefficients, e.g.\ at tree
level from the Wilson term, see below. However, both are odd under
$\mathcal S$ \cref{eq:S}, as are $O_1$ and
$O_8$, while $H_0$ is invariant. Consequently, they do not contribute
at $O(a)$ to energies or to matrix elements of $\mathcal S$-even
operators. This is automatic $O(a)$ improvement in operator form.
The use of $\Phi_1$ and $\Phi_2$ is not even necessary for this
conclusion, since all operators surviving at $\hat m=0$ are
$\mathcal S$-odd.

\subsection{Wilson Clover fermions}

For $\mu_q=0$, $O_8$ and $O_9$ vanish, and after removing $O_1$ with
$\Phi_1$ the operators $O_2,\dots,O_7$ remain. They have the following
interpretation.
\begin{itemize}
\item $O_7$ is absorbed into the improved bare mass
  $\tilde m=\hat m(1+b_m a\hat m)$.
\item $O_5$ and $O_6$ modify the gauge part to
  $\zeta_E\,\mathrm{tr}\,\mathbf E^2+\zeta_B\,\mathrm{tr}\,\mathbf B^2$
  with $\zeta_{E,B}=1+O(a\hat m)$. A canonical rescaling of the gauge
  field shows that this amounts to an improved coupling
  $\tilde g^2=g^2(1+b_g a\hat m)$ and to an $O(a\hat m)$ shift of the
  gauge speed of light.
\item $O_4$ shifts the fermion velocity by $O(a\hat m)$. In the
  Lagrangian picture the corresponding term $\hat m\,\bar\psi\slashed
  D\psi$ is absorbed into the field normalisation, since its temporal
  and spatial parts come with the same coefficient. This is not
  possible in the Hamiltonian, where only the spatial part exists and
  the normalisation of $\psi$ is fixed by the canonical
  anti-commutation relations. A common factor of the fermion and gauge
  velocities only changes the unit of time and is fixed by setting the
  speed of light. $\nu$ in \cref{eq:HWTMgauge}, however, 
  requires an additional improvement coefficient $\nu=\nu_0(1+b_\nu
  a\hat m)$ plus the $O(a^0)$ tuning of $c_0/\nu_0$.
\item $O_2$ and $O_3$ correspond to the Sheikholeslami-Wohlert
  term~\cite{Sheikholeslami:1985ij}. In the isotropic Lagrangian
  formulation they appear in the single combination
  $\bar\psi\sigma_{\mu\nu}F_{\mu\nu}\psi$ with one coefficient
  $c_\mathrm{SW}$. In the Hamiltonian, the magnetic and electric parts
  receive independent coefficients $c_\mathrm{SW}^m$ and
  $c_\mathrm{SW}^e$.
\end{itemize}
Hence, non-perturbative $O(a)$ improvement of the Hamiltonian with
pure Wilson fermions at non-zero quark mass requires two conditions for
$c_\mathrm{SW}^m$ and $c_\mathrm{SW}^e$ and one for $b_\nu$, in
addition to the coefficients $b_m$ and $b_g$ known from the Lagrangian
case. Natural conditions are chiral Ward identities for
$c_\mathrm{SW}^{m,e}$, the mass dependence of the PCAC mass for $b_m$,
and a hadron dispersion relation for $b_\nu$. In a mass-independent
scheme at $\hat m\to0$, only $c_\mathrm{SW}^m$ and $c_\mathrm{SW}^e$
are needed, in addition to the tuning of $c_0/\nu_0$, which is mass independent.

\subsection{Tree level}

At tree level, the Wilson term in \cref{eq:HWTMgauge} yields
$H_1=-\frac r2\,O_1$. Removing $O_1$ with \cref{eq:Phi1} and
$\epsilon=r/4$ gives
\begin{equation}
  H_1^\mathrm{tree}\simeq\frac{rg}{2}\,O_2+\frac{rg}{4}\,O_3+\frac{r}{2}\,O_4\,.
\end{equation}
The improved Hamiltonian therefore contains the lattice clover
counterterm
\begin{equation}
  -\frac{\nu arg}{2}\sum_{\mathbf n}\psi^\dagger_{\mathbf n}
  \left[c_\mathrm{SW}^m\,\beta\,\Sigma_k\hat B_{\mathbf n,k}
  +c_\mathrm{SW}^e\,i\beta\,\alpha_k\hat E_{\mathbf n,k}\right]\psi_{\mathbf n}\,.
\end{equation}
Here
\begin{equation}
  \hat E^a_{\mathbf n,k}=\frac{g}{2a^2}(R^a_{\mathbf n-\mathbf e_k,k}
  - L^a_{\mathbf n,k})\,,\quad \hat E_{\mathbf n,k} = T^a \hat E^a_{\mathbf n,k}
\end{equation}
is the site-centred electric field, the symmetric combination of the
generators acting at site $\mathbf{n}$. This symmetric combination is 
required by invariance under $\mathcal P_\mathrm{tm}$.
$\hat
B_{\mathbf n,k}^a$ can be obtained like in the Euclidean case from
the clover average over the plaquettes connected to site $\mathbf n$
in the corresponding plane. The normalisation of the clover
counterterm is such that the isotropic Euclidean action corresponds to
$c_\mathrm{SW}^m=c_\mathrm{SW}^e=c_\mathrm{SW}$. The tree level values
for $c_\mathrm{SW}^{m,e}$ and $b_\nu$ can be read out from above
\begin{equation}
  c_\mathrm{SW}^m=1\,,\qquad c_\mathrm{SW}^e=\frac12\,,\qquad
  b_\nu=-\frac r2\,,\qquad \nu_0=1
\end{equation}
in the basis where $O_1$ has been eliminated.
The electric coefficient $c_\mathrm{SW}^e=c_\mathrm{SW}^m/2$
because the temporal Wilson term is absent in the Hamiltonian. This
agrees with the an-isotropic clover action of
Ref.~\cite{Harada:2001ei}, see also Ref.~\cite{Umeda:2000ym}.

\section{Free Energies}

As a cross-check, we compute the lattice artefacts in the
energies of the free Wilson twisted mass fermion Hamiltonian
\begin{equation}
  \label{eq:HWTM}
  \begin{aligned}
    H_{\text{WTM}}
    &=
    m_q
    \sum_{\mathbf n}
    \Psi_{\mathbf n}^\dagger
    \beta
    \Psi_{\mathbf n}
    +
    i\mu_q
    \sum_{\mathbf n}
    \Psi_{\mathbf n}^\dagger
    \beta\gamma_5\tau_3
    \Psi_{\mathbf n}
    \\[1ex]
    &\quad
    -
    \frac{i}{2a}
    \sum_{\mathbf n,i}
    \Psi_{\mathbf n}^\dagger
    \alpha_i
    \left(
    \Psi_{\mathbf n+\mathbf e_i}
    -
    \Psi_{\mathbf n-\mathbf e_i}
    \right)
    \\[1ex]
    &\quad
    +
    \frac{r}{2a}
    \sum_{\mathbf n,i}
    \Psi_{\mathbf n}^\dagger
    \beta
    \left(
    2\Psi_{\mathbf n}
    -
    \Psi_{\mathbf n+\mathbf e_i}
    -
    \Psi_{\mathbf n-\mathbf e_i}
    \right)
  \end{aligned}
\end{equation}
in $d=3$ spatial dimensions. In $d=1$ the corresponding result can be
found in Ref.~\cite{Schwagerl:2025dsj}.
Let us introduce $\hat p_i(\mathbf{p}) = \sin(ap_i)/a$, as well as
the Wilson term in momentum space
\begin{equation}
  W(\mathbf p)
  =
  \frac{r}{a}
  \sum_{i=1}^3
  \left[1-\cos(p_i a)\right]
  =
  \frac{2r}{a}
  \sum_{i=1}^3
  \sin^2\left(\frac{p_i a}{2}\right)\,.
\end{equation}
With these definitions, we obtain the single-particle Hamiltonian for
momentum $\mathbf{p}$
\begin{equation}
  h(\mathbf p)
  =
  \beta\left[m_q+W(\mathbf p)\right]
  +
  i\mu_q\,\beta\gamma_5\tau_3
  +
  \sum_{i=1}^3
  \alpha_i\,\hat{p}_i(\mathbf p).
\end{equation}
after Fourier transforming the Hamiltonian from above.
Setting $m_W(\mathbf{p}) = m_q + W(\mathbf{p})$ we may define
\begin{equation}
  \mathcal M(\mathbf p)
  =
  \sqrt{
    m_W(\mathbf p)^2+\mu_q^2
  }\,.
\end{equation}
Using the commutation relations of the $\gamma$-matrices, it is easy
to see that
\begin{equation}
  h(\mathbf p)^2
  =
  \left[
    \mathcal M(\mathbf p)^2
    +
    \sum_{i=1}^3
    \hat{p}_i(\mathbf p)^2
    \right]\mathbf 1\,,
\end{equation}
and thus the energy eigenvalues read
\begin{equation}
  E_\pm(\mathbf p)
  =
  \pm
  \sqrt{
    \left[m_q+W(\mathbf p)\right]^2
    +
    \mu_q^2
    +
    \sum_{i=1}^3
    \hat{p}_i(\mathbf p)^2
  }\,.
\end{equation}
Since $W(\mathbf 0)=0$ and $\hat{p}_i(\mathbf 0)=0$, we have the
special case
\begin{equation}
  E_\pm(\mathbf 0)
  =
  \pm\mathcal M_q
  =
  \pm\sqrt{m_q^2+\mu_q^2}\,.
\end{equation}
From this one can conclude
\begin{equation}
  b_m=0
\end{equation}
at tree level, unlike $b_m=-\frac12$ in the Euclidean case, where it
stems from the temporal discretisation.

\subsection{Lattice artefacts in $E_\pm$}

For $a\ll 1$ we may expand
\begin{equation}
  W(\mathbf p)
  =
  \frac{r a}{2}\mathbf p^2
  +
  O(a^3)\,,
\end{equation}
and
\begin{equation}
  \sum_i \hat{p}_i(\mathbf p)^2
  =
  \mathbf p^2
  -
  \frac{a^2}{3}
  \sum_i p_i^4
  +
  O(a^4)\,.
\end{equation}
Defining $E_{\mathrm{cont}}(\mathbf{p}) = +\sqrt{\mathcal M_q^2+\mathbf p^2}$, one
obtains the following expansion for $E_\pm$
\begin{equation}
  \label{eq:Eexpansion}
  \begin{split}
    E_\pm(\mathbf p)
    =&
    \pm
    E_\mathrm{cont}
    \pm am_q r
    \frac{\mathbf p^2}{2E_{\text{cont}}}\\
    &\pm
    a^2
    \frac{
      \dfrac{r^2}{4}\mathbf p^4
      -
      \dfrac{1}{3}\sum_i p_i^4
    }
         {
           2E_\mathrm{cont}
         } \mp a^2\frac{m_q^2r^2\mathbf{p}^4}{8E_\mathrm{cont}^3}\\
         &+
         O(m_qa^3, a^4)\,.\\
  \end{split}
\end{equation}
Note that the term proportional to $\sum_i p_i^4$ is absent in $d=1$
dimension. The $O(m_q a^3)$ correction reads explicitly 
\begin{equation}
  \begin{split}
    a^3m_qr\left(\mp \frac{1}{24E_\mathrm{cont}}\sum_i p_i^4 \right. &\mp
    \frac{\mathbf{p}^2}{16E_\mathrm{cont}^3}\left(r^2\mathbf{p}^4-\frac{4}{3}\sum_ip_i^4\right)\\
    &\pm \left.\frac{m_q^2r^2\mathbf{p}^6}{16E_\mathrm{cont}^5}\right)\,.\\
  \end{split}
\end{equation}
This result is consistent with automatic $O(a)$ improvement proven
above, as the artefacts odd in $a$ vanish at $m_q=0$. Moreover,
$a$-odd corrections are odd in both $m_q$ and $r$, which is leading
to $O(a)$ improvement via mass and Wilson average, respectively.

The tree-level value of $b_\nu$ is confirmed by the exact free dispersion
relation for the pure Wilson case $E^2=(m_q+W(\mathbf
p))^2+\sum_i\hat p_i^2\simeq m_q^2+\mathbf p^2(1+ram_q)$.

\section{Conclusion}

In this paper we have applied the framework of the Symanzik effective
Hamiltonian proposed in Ref.~\cite{Steffens:2026doc} to Wilson twisted
mass fermions. In contrast to pure gauge theory, artefacts linear in
the lattice spacing appear. We have shown that automatic $O(a)$
improvement is in place 
in the Hamiltonian formalism of Wilson twisted mass fermions at
maximal twist. In fact, at maximal twist all contributions to
$\mathcal S$-even observables odd in the lattice spacing are absent by 
symmetry arguments. Likewise, what is called mass and Wilson averages lead to
the absence of odd powers of the lattice spacing in $\mathcal S$-even
quantities.

Moreover, we have explicitly derived the 
list of contributions to the effective Hamiltonian at $O(a)$ of
dimension-5. We have shown how to use unitary transformations to
remove redundant operators from this list for on-shell
improvement. The remaining contributions 
include the well known Sheikholeslami-Wohlert term, which splits into
two terms in the Hamiltonian formulation due to missing boosts.
We have derived the tree-level values for the
corresponding two independent coefficients, with values consistent to
results from the an-isotropic lattice clover action. 

Given appropriate non-perturbative improvement conditions, this result
can be used to define a non-perturbative improvement programme also in
the Hamiltonian formalism both for on-shell and operator improvement.
It is worth noting, however, that the number of parameters to tune for
on-shell improvement increases due to reduced Lorentz symmetry
compared to the isotropic space-time approach.

\begin{acknowledgments}
  The author thanks Marco Garofalo and Fernanda Steffens for helpful
  comments to the manuscript.
  
  This project was funded by the Deutsche Forschungsgemeinschaft (DFG,
  German Research Foundation) as a project in the CRC 1639 NuMeriQS --
  project no.\ 511713970, and under Germany’s Excellence Strategy –
  EXC 3107 – Project-ID 533766364 in the Color-meets-flavor cluster of
  excellence.

  We have used Claude AI for checking the consistency of the
  manuscript and for performing some of the basic calculations.
\end{acknowledgments}

\appendix
\section{One-derivative bilinears in $H_1$}
\label{app:onederiv}

For completeness, we provide here more details on dimension-5
densities with a single covariant derivative of the form
$\lambda\,\Psi^\dagger M_i\,(-iD_i)\,\Psi$ with $\lambda\in\{\hat
m,\mu_q\}$ and a Hermitian matrix $M_i$ in spinor and                                               
flavour space. 
Up to normalisation, the elements of the Clifford algebra with one spatial
index are $\gamma^0\gamma^i=\alpha_i$, $\gamma^i=\beta\alpha_i$,
$\gamma^0\gamma^i\gamma_5=\Sigma_i$ and
$\gamma^i\gamma_5=\beta\Sigma_i$. Also,
$\sigma^{0i}$ and $\epsilon_{ijk}\sigma^{jk}$ are proportional to $\alpha_i$
and $\Sigma_i$, respectively. Exact $U(1)_3$ invariance restricts the flavour
structure to $\mathbf{1}$ and $\tau_3$. With the Hermitian choices
$M_i\in\{\alpha_i,\, i\beta\alpha_i,\, \Sigma_i,\, \beta\Sigma_i\}\otimes
\{\mathbf{1},\tau_3\}$ this gives 16 candidates in total of this form. Under
$\mathcal{P}_\mathrm{tm}$ we use $D_i\to-D_i$, under the anti-unitary
$\mathcal{T}_\mathrm{tm}$ the complex conjugation acts on $M_i$ and on
$-iD_i$, under $\mathcal{C}$ the transposition includes a partial
integration (since $\mathcal{C}$ moves the derivative onto
$\Psi^\dagger$), and under $\tau_1$ the twisted mass flips sign, 
$\mu_q\to-\mu_q$. \Cref{tab:onederiv} lists the symmetries under which
each candidate is odd. The $\Theta$ entries equal the products of the
$\mathcal{C}$, $\mathcal{P}_\mathrm{tm}$ and $\mathcal{T}_\mathrm{tm}$
signs, and are consistent with \cref{eq:chik}. The only candidate surviving all
constraints is $O_4=\hat m\,\psi^\dagger\alpha_i(-iD_i)\psi$.

\begin{table}[h]
\centering
\begin{tabular}{l|cccc}
 $M_i$ & $\hat m\,\mathbf{1}$ & $\hat m\,\tau_3$ & $\mu_q\,\mathbf{1}$ & $\mu_q\,\tau_3$\\
\hline
$\alpha_i$            & allowed ($O_4$)                      & $\mathcal{P}_\mathrm{tm},\mathcal{T}_\mathrm{tm},\tau_1$ & $\tau_1$                                  & $\mathcal{P}_\mathrm{tm},\mathcal{T}_\mathrm{tm}$\\
$i\beta\alpha_i$      & $\mathcal{T}_\mathrm{tm},\Theta$     & $\mathcal{P}_\mathrm{tm},\Theta,\tau_1$                   & $\mathcal{T}_\mathrm{tm},\Theta,\tau_1$   & $\mathcal{P}_\mathrm{tm},\Theta$\\
$\Sigma_i$            & $\mathcal{P}_\mathrm{tm},\mathcal{C}$ & $\mathcal{T}_\mathrm{tm},\mathcal{C},\tau_1$             & $\mathcal{P}_\mathrm{tm},\mathcal{C},\tau_1$ & $\mathcal{T}_\mathrm{tm},\mathcal{C}$\\
$\beta\Sigma_i$       & $\mathcal{P}_\mathrm{tm},\Theta$     & $\mathcal{T}_\mathrm{tm},\Theta,\tau_1$                   & $\mathcal{P}_\mathrm{tm},\Theta,\tau_1$   & $\mathcal{T}_\mathrm{tm},\Theta$\\
\end{tabular}
\caption{Mass-weighted one-derivative bilinears $\lambda\,\psi^\dagger M_i(-iD_i)\psi$
and the symmetries under which they are odd.}
\label{tab:onederiv}
\end{table}
 
\end{document}